\documentclass[preprint]{ptephy_v1}

\preprintnumber{YITP-16-71} 

\usepackage{epstopdf}
\usepackage{graphicx}

\begin{document}

\title{Creating and probing the Sachdev-Ye-Kitaev model with ultracold gases: 
\\ 
Towards experimental studies of quantum gravity}

\author{Ippei Danshita$^{1,*}$}
\author{Masanori Hanada$^{1,2,3}$}
\author{Masaki Tezuka$^{4}$}
\affil{
{$^1$Yukawa Institute for Theoretical Physics, Kyoto University, Kyoto 606-8502, Japan}
\\
$^2$Stanford Institute for Theoretical Physics, Stanford University, Stanford, CA 94305, USA
\\
$^3$The Hakubi Center for Advanced Research, Kyoto University, Kyoto 606-8501, Japan
\\
{$^4$Department of Physics, Kyoto University, Kyoto 606-8502, Japan}
\\
{$^*$danshita@yukawa.kyoto-u.ac.jp}
}



\begin{abstract}%
We suggest that the holographic principle, combined with recent technological advances in atomic, molecular, and optical physics, can lead to experimental studies of quantum gravity. 
As a specific example, we consider the Sachdev-Ye-Kitaev (SYK) model, which consists of spin-polarized fermions with an all-to-all complex random two-body hopping and has been conjectured to be dual to a certain quantum gravitational system. Achieving low-temperature states of the SYK model is interpreted as a realization of a stringy black hole, provided that the holographic duality is true. 
We introduce a variant of the SYK model, in which the random two-body hopping is real. This model is equivalent to the origincal SYK model in the large-$N$ limit. We show that this model can be created in principle
by confining ultracold fermionic atoms into optical lattices and coupling two atoms with molecular states via photo-association lasers. This development serves as an important first step towards an experimental realization of such systems dual to quantum black holes.
We also show how to measure out-of-time-order correlation functions of the SYK model, which allow for identifying the maximally chaotic property of the black hole.
\end{abstract}

\subjectindex{I22, A63, B21}

\maketitle

\section{Introduction}
The quantum nature of black holes is one of the most important subjects in theoretical physics, since the theoretical discovery of particle-emissions from a black hole due to quantum effects~\cite{Hawking:1974rv,Hawking:1974sw}, which are often referred to as the Hawking radiation. Although there have been experimental searches for quantum black holes at the CERN LHC motivated by the predictions on the basis of theories of TeV-scale quantum gravity~\cite{dimopoulos-01,giddings-02,calmet-10}, no evidence of the black hole creation has been observed thus far~\cite{cms-13, cms-15, cms-16, atlas-14}. In this paper, we present a completely different route to experimental studies of quantum gravity by exploiting both holographic principle and unprecedented controllability of optical-lattice systems loaded with ultracold gases~\cite{footnote1}.

In order to resolve paradoxes associated with the black hole evaporation that results from the Hawking radiation, the holographic principle \cite{'tHooft:1993gx,Susskind:1994vu} emerged, 
which claims that black holes, and more general quantum gravitational theories, are equivalent to non-gravitational theories in different spacetime dimensions. 
As a concrete example, the gauge/gravity duality conjecture \cite{Maldacena:1997re},  
which claims the duality (i.e. the equivalence) between superstring/M-theory on certain spacetimes 
and quantum field theories, has been studied extensively. Although this conjecture has not been proven yet, it is believed to be correct at least in some simplest cases. 
For example, maximally supersymmetric matrix quantum mechanics (also known as the Matrix Model of M-theory \cite{Banks:1996vh,deWit:1988ig}), 
which is conjectured to describe a black hole in type IIA superstring theory 
near the 't Hooft large-$N$ limit \cite{Itzhaki:1998dd}, has been studied numerically starting in \cite{Anagnostopoulos:2007fw}.
The agreement with the dual superstring theory prediction has been confirmed 
including the effect of virtual loops of string \cite{Hanada:2013rga}.

Thanks to their high controllability and cleanness, experiments with ultracold gases in optical lattices have succeeded in realizing various theoretical models, which were introduced in the contexts of condensed matter physics but did not have quantitative experimental counterparts. Examples include the Bose-Hubbard model~\cite{greiner-02}, the Lieb-Liniger model~\cite{moritz-03, kinoshita-04}, the Aubry-Andr\'e model~\cite{roati-08}, the Harper Hamiltonian~\cite{aidelsburger-13,miyake-13}, and the topological Haldane model~\cite{jotzu-14}. There have been theoretical proposals also for realizing lattice gauge models studied in high-energy physics~\cite{kasamatsu-13,Wiese:2014rla,Zohar:2015hwa}. These circumstances tempt one to expect that it may be possible as well to realize quantum field theories dual to quantum gravitational systems.

In this paper, we propose a possible way to create the Sachdev-Ye-Kitaev (SYK) model~\cite{Sachdev:2015efa, Kitaev_talk, maldacena-16, jensen-16} experimentally with use of ultracold gases in optical lattices. The SYK model consists of spin-polarized fermions with an all-to-all random two-body hopping. Its thermal state is a non-Fermi liquid with nonzero entropy at vanishing temperature, which is called the Sachdev-Ye (SY) state~\cite{sachdev-93}, and has been conjectured to be holographically dual to charged black holes with two-dimensional anti-de Sitter (AdS$_2$) horizons~\cite{Sachdev:2015efa,sachdev-10}. For the purpose of experimental realization, this model is advantageous over the other known models with holography in the sense that it consists of non-relativistic particles and is not supersymmetric. Here we emphasize that the experimental realization of the SY state in optical-lattice systems is equivalent to that of a quantum black hole if the duality is true. 

Our strategy to achieve the SYK model is twofold. We first simplify the model into a form that can be accessed more easily in experiments. Specifically, we numerically demonstrate that the original SYK model, which has a complex two-body hopping with Gaussian randomness, can be quantitatively approximated by the model possessing a real two-body hopping mediated via random couplings to multiple molecular states. The SYK model is exactly reproduced in the limit with infinitely many molecular states.
Second, we show that the latter model can be created in principle by confining ultracold fermionic atoms into a deep optical lattice and utilizing photo-association (PA) lasers~\cite{jones-06} that couple all the combinations of two atomic bands with molecular states. However, a practical realization of the proposed scheme is still difficult even with current experimental technology. We describe such practical difficulties together with possible solutions to some of them.
We also present a protocol to measure two physical quantities characterizing the black hole dual to the SY state, namely out-of-time-order correlation (OTOC) functions~\cite{Kitaev_talk,Maldacena:2015waa} and single-particle Green's function~\cite{Sachdev:2015efa}.
In the following, we set the reduced Planck's constant and the Boltzmann constant to be $\hbar = k_{\rm B} = 1$.

\section{Sachdev-Ye-Kitaev model}
\label{sec:SYK}
The SYK model~\cite{Sachdev:2015efa,Kitaev_talk} is a model of $Q$ spin-polarized fermions on $N$ sites.
The Hamiltonian is given by~\cite{footnote2}
\begin{eqnarray}
\hat{H}=\frac{1}{(2N)^{3/2}}\sum_{ijkl}J_{ij,kl}\hat{c}^\dagger_i\hat{c}^\dagger_j\hat{c}_k\hat{c}_l, \label{Hamiltonian_SYK}
\end{eqnarray}
where indices run from $1$ to $N$, the creation and annihilation operators $\hat{c}^\dagger_i$ and $\hat{c}_i$ satisfy the anti-commutation relations
\begin{eqnarray}
\{\hat{c}_i,\hat{c}_j\}=\{\hat{c}^\dagger_i,\hat{c}^\dagger_j\}=0, 
\qquad
\{\hat{c}^\dagger_i,\hat{c}_j\}=\delta_{ij}, 
\end{eqnarray}
and $J_{ij,kl}$ is a complex Gaussian random coupling constant which satisfies
\begin{eqnarray}
J_{ij,kl}
=
-J_{ji,kl}
=
-J_{ij,lk}, \,\,
J_{ij,kl}
=
J_{kl,ij}^\ast,
\end{eqnarray}
and
\begin{eqnarray}
\overline{({\rm Re}~J_{ij,kl})^2}=\begin{cases}J^2/2&(\{i,j\}\neq\{k,l\})\\J^2&(\{i,j\}=\{k,l\})\end{cases},
\\
\overline{({\rm Im}~J_{ij,kl})^2}=\begin{cases}J^2/2&(\{i,j\}\neq\{k,l\})\\0&(\{i,j\}=\{k,l\})\end{cases}.
\end{eqnarray}
Here $\overline{\ \cdot\ }$ stands for the disorder average.  
This system is strongly coupled when $J/T$ ($T$: temperature) is large. 
Only planar diagrams survive in $N\to\infty$  with $J$ fixed.  
In the following we take $J$ as the unit of energy.

This system, in the large-$N$ and strong coupling limit, has properties strikingly resembling a black hole. 
Firstly, Sachdev \cite{Sachdev:2015efa} pointed out that this theory has the same entropy density as a black hole in AdS$_2$. 
He also found the agreement of several correlation functions. 
Furthermore, Kitaev \cite{Kitaev_talk} calculated the Lyapunov exponent and found that it has a pattern proposed by Maldacena {\it et al.}~\cite{Maldacena:2015waa} 
for quantum theories with dual gravity description. Namely, the Lyapunov exponent 
takes the maximum value $2\pi T$ at strong coupling limit $J/T\to\infty$. 
Therefore it has been expected that the SYK model is actually equivalent to classical gravity in the large-$N$ limit.
Then, because this theory admits the $1/N$-expansion,
it is natural to expect that $1/N$ correction describes the effect of loops of strings in a similar way to the case of gauge theories \cite{'tHooft:1973jz,footnote3}.

We slightly modify the SYK model in order to make the experimental implementation easier. 
The Hamiltonian is still \eqref{Hamiltonian_SYK}, but the random coupling $J_{ij,kl}$ is taken to be real. 
The Gaussian random coupling is modified to
\begin{eqnarray}
J_{ij,kl}
=
-J_{ji,kl}
=
-J_{ij,lk}, 
\end{eqnarray}
\begin{eqnarray}
J_{ij,kl}
=
J_{kl,ij}
\end{eqnarray}
\begin{eqnarray}
\overline{|J_{ij,kl}|^2}
=\begin{cases}J^2&(\{i,j\}\neq\{k,l\})\\2J^2&(\{i,j\}=\{k,l\})\end{cases},
\end{eqnarray}
and for $\{i,j\}\neq\{k,l\}$
\begin{eqnarray} 
\overline{J_{ij,kl}J_{pq,rs}}=J^2\left\{
\left(\delta_{ir}\delta_{js}-\delta_{is}\delta_{jr}\right)
\left(\delta_{kp}\delta_{lq}-\delta_{kq}\delta_{lp}\right) \right.
\nonumber \\ 
+\left.
\left(\delta_{ip}\delta_{jq}-\delta_{iq}\delta_{jp}\right)
\left(\delta_{kr}\delta_{ls}-\delta_{ks}\delta_{lr}\right)
\right\}.  
\label{eq:realJJave}
\end{eqnarray} 
Coefficients have been chosen so that the eigenenergy distribution coincides with that of the original SYK model.
The second term inside $\{\cdots\}$ in \eqref{eq:realJJave} is absent in the original SYK model. Due to this, each Feynman diagram receives some correction 
after disorder average.
However, such corrections are $1/N$-suppressed in general, 
and hence this modified model agrees with the original model at large-$N$.
In the following, we call the original SYK model with complex $J_{ij,kl}$ and the modified one with real $J_{ij,kl}$ 
`complex-SYK' and `real-SYK,' respectively. 
The $1/N$-corrections to the real- and complex-SYK models are described by different sets of Feynman diagrams. 
In analogy to the duality between gauge theory and superstring, it is natural to expect that these two theories describes slightly different 
quantum gravitational systems whose classical limits coincide.
In Appendix A, we indeed perform numerical comparisons between the two models to demonstrate that the computed physical quantities of the two models rapidly approaches each other as $N$ increases.

\begin{figure}[tb]
\centering
     \includegraphics[scale=0.73]{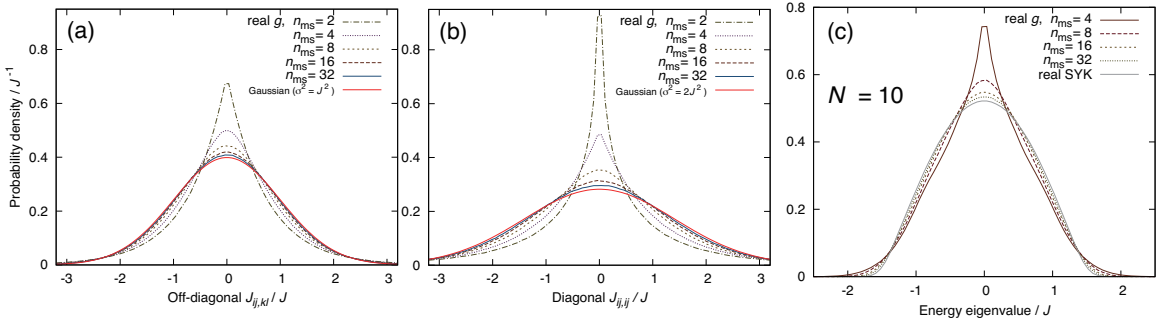}
   \caption{(a): Distribution of $J_{ij,kl}=\frac{(2N)^{3/2}}{\sqrt{n_{\rm ms}}J}\left(\sum_{s:{\rm even}}g_{s,ij}g_{s,kl}-\sum_{s:{\rm odd}}g_{s,ij}g_{s,kl}\right)$ with only the off-diagonal components (i.e. $(i,j)\neq (k,l),(l,k)$);
(b): Distribution of $J_{ij,ij}=\frac{(2N)^{3/2}}{\sqrt{n_{\rm ms}}J}\left(\sum_{s:{\rm even}}g_{s,ij}^2-\sum_{s:{\rm odd}}g_{s,ij}^2\right)$.
The numbers of samples taken are $10^4$ (a) and $10^5$ (b), respectively. 
(c): The energy spectrum for $N=10$, $Q=N/2$, and $10^4$ samples. 
For all of (a), (b), and (c), the weight of real $g_{s,ij}$ is Gaussian, $\frac{e^{-g_{s,ij}^2/(2\sigma_g^2)}}{\sqrt{2\pi}\sigma_g}$ with $\sigma_g^2 = (2N)^{-3}J^2$ while $\nu_s=+\sqrt{n_{\rm ms}}J$ for even $s$ and $\nu_s=-\sqrt{n_{\rm ms}}J$ for odd $s$.
   }
\label{fig:dist_gg_N10_real}
\end{figure} 

One of the severest bottlenecks for realizing the SYK model in optical-lattice experiments is the implementation of the all-to-all two-body hopping, because particles on lattice systems in general move the most dominantly via nearest-neighbor one-body hopping. In order to overcome this bottleneck, we consider a situation, in which two atoms are coupled with $n_{\rm ms}$ molecular states, described by the following Hamiltoinan,
\begin{eqnarray}
\hat{H}_{\rm m}
\! =\!
\sum_{s=1}^{n_{\rm ms}}\! \left\{\nu_s \hat{m}_s^\dagger\hat{m}_s
+ \sum_{s'=1}^{n_{\rm ms}} \frac{U_{s,s'}}{2} 
\hat{m}_s^{\dagger} \hat{m}_{s'}^{\dagger} \hat{m}_{s'} \hat{m}_{s}
\!+\!
\sum_{i,j}g_{s,ij}\!
\left(
\hat{m}_s^\dagger\hat{c}_i\hat{c}_j
-
\hat{m}_s\hat{c}_i^\dagger\hat{c}_j^\dagger
\right)\!
\right\}.  
\label{Hamiltonian_multi_molecular_states}
\end{eqnarray}
Here, $\nu_{s}$, $U_{s,s'}$, and $g_{s,ij}$ denote the detuning of molecular state $s$, the onsite interaction between two molecules in states $s$ and $s'$, and the atom-molecule coupling constant. Using the degenerate perturbation theory up to the second order, we obtain the following effective Hamiltonian,
\begin{eqnarray}
\hat{H}_{\rm eff}
=
\sum_{s,i,j,k,l}\frac{g_{s,ij}g_{s,kl}}{\nu_s}\hat{c}^\dagger_i\hat{c}^\dagger_j\hat{c}_k\hat{c}_l. 
\label{Hamiltonian_AMO}
\end{eqnarray}
See Appendix~C for more detailed derivations of the effective Hamiltonian.
A similar way of designing a kind of two-body hopping term, namely the ring exchange interaction, by means of intermediate two-particle states has been pointed out in previous work~\cite{Buchler:2005fq}. 
In the next section we elaborate how to prepare such a situation in optical-lattice systems while in this section we show that Eq.~(\ref{Hamiltonian_AMO}) serves as a quantitative approximation of the complex SYK model (\ref{Hamiltonian_SYK}) when $n_{\rm ms}$ is sufficiently large and $\nu_s$ is appropriately tuned. 

Let us suppose $\nu_1=\nu_2=\cdots=\nu_{n_{\rm ms}}\propto \sqrt{n_{\rm ms}}$. Then, if $n_{\rm ms}$ is large enough, $\sum_{s}\frac{g_{s,ij}g_{s,kl}}{\nu_s}$ should become Gaussian except for the diagonal elements $(i,j)=(k,l)$ or $(i,j)=(l,k)$ (note that $g_{s,ij}^2$ is always positive). This happens because it is simply an $n_{\rm ms}$-step random walk for each set of indices $(i,j,k,l)$.
In order to improve the behavior of the diagonal elements, we take $n_{\rm ms}$ to be even, 
and set $\nu_s = +\sqrt{n_{\rm ms}} \sigma_{\nu}$ for even $s$ and $\nu_s = -\sqrt{n_{\rm ms}} \sigma_{\nu}$ for odd $s$. 
We assume that the distribution of the real $g_{s,ij}$ is Gaussian having the variance $\sigma^2 = \sigma_g^2$,
with $\sigma_g^2 / \sigma_{\nu} = J/(2N)^{3/2}$.
In this section we set $\sigma_{\nu} =\sigma_{\rm g}= J/(2N)^{3/2}$ for simplicity.

As explained in Appendix B, if we identify $\sum_{s}\frac{g_{s,ij}g_{s,kl}}{\nu_s}$
defined in this way with $J_{ij,kl}/(2N)^{3/2}$, 
the properties needed in the real-SYK model are satisfied at $n_{\rm ms}=\infty$. 
We collected samples by using independent real Gaussian random values of $\{g_{ij}\}$.
In Fig.~\ref{fig:dist_gg_N10_real}, we plot the distribution of $J_{ij,kl}$ with $\{i,j\}\neq \{k,l\}$ and $10^4$ samples (a) and the diagonal elements $J_{ij,ij}$ with $10^5$ samples (b).
The distributions have different shapes for smaller values of $n_{\rm ms}$, but they quickly approach Gaussian distributions with corresponding variances for the real-SYK model as $n_{\rm ms}$ increases.
In Fig.~\ref{fig:dist_gg_N10_real}(c), we plot the energy spectrum of this model using $10^4$ samples with $N=10$ and $Q=N/2$. 
The energy spectra become closer to that of the real-SYK model as $n_{\rm ms}$ increases (for comparisons regarding other quantities, see Appendix A).

\section{Creating the model}
\label{sec:optical_lattice}

\begin{figure}[tb]
\centering
\includegraphics[scale=0.6]{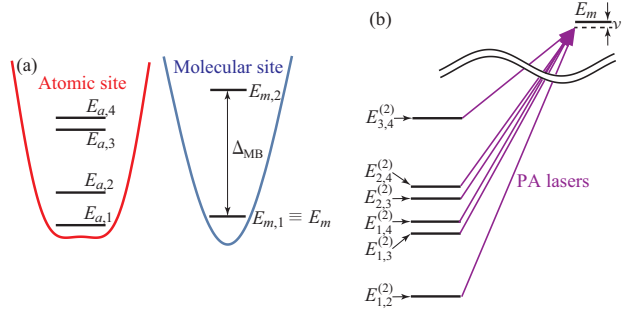}
\caption{\label{fig:levels}
Schematic illustrations of the energy levels of the atomic and molecular states relevant to our protocol (a) and the PA process (b) for $N=4$ and $n_{\rm ms} = 1$.
}
\end{figure}

In this section, we explain how to create the model (\ref{Hamiltonian_AMO}), a simplified version of the SYK model, in a system of optical lattices loaded with ultracold gases. We consider a two-dimensional gas of spin-polarized fermionic atoms confined in an optical lattice. 
In the proposed scheme, we utilize the PA process that coherently converts two atoms into a bosonic molecule in a certain electronic (or hyperfine), vibrational, and rotational state~\cite{jones-06}. 
We assume that molecules are confined also by the optical-lattice lasers confining atoms.
However, since in general the lattice depth for molecules may be controlled independently from that for atoms, we assume that the former has the sign opposite to the latter. 
In this situation, the potential minima of the molecular optical lattice sit right next to those of the atomic optical lattice, as illustrated in Fig.~\ref{fig:levels}(a), such that we do not have to take into account the effects of the onsite interactions between an atom and a molecule, which would otherwise complicate the levels of the atomic and molecular bands. 
We assume that the optical lattices are so deep that atoms and molecules in each lattice site are completely isolated. To make the manipulation of the system easier, we remove all the atoms in the lattice sites neighboring to occupied sites. We also assume that each occupied atomic lattice site contains $Q$ atoms. We regard the band degrees of freedom in the atomic site as the physical site index of the SYK model. More specifically, the first, second, third, $\ldots$, $N$-th bands correspond to $i=1, 2, 3, \ldots, N$ sites. We write the energy of the lowest molecular band and that of the $i$-th atomic band as $E_m$ and $E_{a,i}$. 

Let us introduce a PA laser, which couples atomic bands $i(\leq N)$ and $j (\leq N)$ with the lowest molecular band. The frequency of the PA laser is chosen as
\begin{eqnarray}
\omega^{\rm PA}_{i,j} = E_m - E^{(2)}_{i,j} - \nu,
\end{eqnarray}
where $E^{(2)}_{i,j}= E_{a,i} + E_{a,j}$, and $\nu$ denotes the detuning. We consider a situation in which all the combinations of the two atomic bands $(i,j)$ are coupled via independent PA lasers as shown in Fig.~\ref{fig:levels}(b). For such a situation to be possible, $|\nu|$ has to be larger than the linewidth of the PA lasers $\Gamma_{\rm PA}$ and that of the molecular state $\Gamma_{\rm ms}$. In addition, the condition $|\nu|\ll \Delta_{\rm min}$ has to be satisfied, where $\Delta_{\rm min}$ denotes the minimum level spacing in $E^{(2)}_{i,j}\leq E^{(2)}_{N-1,N}$. The number of necessary PA lasers is $N(N-1)/2$.
The PA process is described by the following Hamiltonian, 
\begin{eqnarray}
\hat{H}_{\rm m1} = \nu \hat{m}^{\dagger}\hat{m}
+ \frac{U}{2}\hat{m}^{\dagger}\hat{m}^{\dagger}\hat{m}\hat{m}
+ \sum_{i,j}g_{ij} (\hat{m}^{\dagger}\hat{c}_{j}\hat{c}_{i}+ h.c.),
\end{eqnarray}
where the atom-molecule coupling constant is given by
\begin{eqnarray}
g_{ij} \!=\! \frac{1}{2}{\rm sgn}(j-i) \!
\int \! d{\boldsymbol r}\, \Omega_{i,j}({\bf r})
w_{m}({\boldsymbol r})
w_{a,i}\left({\boldsymbol r}\right)
w_{a,j}\left({\boldsymbol r}\right).
\end{eqnarray}
$\Omega_{i,j}({\boldsymbol r})$ denotes the intensity of the PA laser while $w_{m}({\boldsymbol r})$ and $w_{a,i}({\boldsymbol r})$ represent the Wannier function of the 1st molecular band and the $i$-th atomic band. The absolute value of the detuning $|\nu|$ is assumed to be much smaller than the onsite interaction $U$ between two molecules in order to avoid double occupancy of the molecules. For the same reason, $U$ has to be much smaller than the minimum level spacing $\Delta_{\rm min}$ in $E^{(2)}_{i,j}$ or sufficiently larger than the maximum level spacing $\Delta_{\rm max}$ in $E^{(2)}_{i,j}$. Moreover, the level spacing between the first and second molecular bands $\Delta_{\rm MB}$ is assumed to be larger than $\Delta_{\rm max}$ in order to avoid accidental couplings between higher molecular bands and the atomic bands via the PA lasers.

\begin{figure}[tb]
\centering
\includegraphics[scale=0.5]{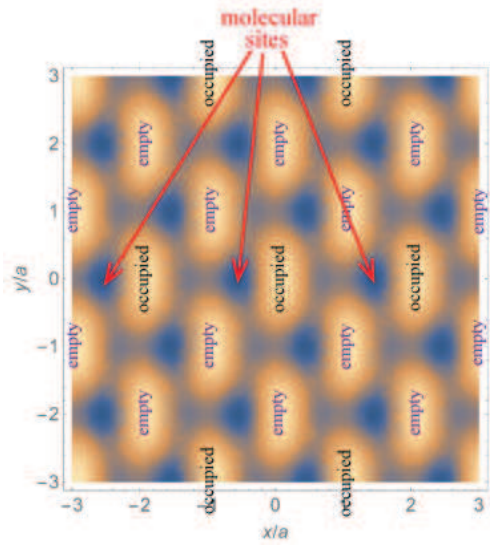}
\caption{\label{fig:dwl}
Spatial profile of the optical lattice of Eq.~(\ref{eq:DWL}) for $V_0<0$, $R=0.59$, and $\theta = \pi/6$. The dark and light colors indicate the high- and low-potential regions. This means that the lightest (darkest) spots correspond to the atomic (molecular) sites.
}
\end{figure}

A PA molecule has many vibrational and rotational states. When $\Delta_{\rm max} < \tilde{\Delta}$, we can extend the scheme described above to include couplings of two atoms with multiple molecular states, where $\tilde{\Delta}$ denotes the minimum level spacing of the involved molecular states. The extended system is now described by Eq.~(\ref{Hamiltonian_multi_molecular_states}).
When $|\nu_{s}|\gg |g_{s,ij}|$, one can integrate out the molecular degrees of freedom through the second-order perturbation theory with respect to the atom-molecule couplings, leading to the effective Hamiltonian of Eq.~(\ref{Hamiltonian_AMO}). Notice that the precise condition for the second-order perturbation theory to be valid is shown in Appendix C.
We emphasize that the coupling constant $g_{s,ij}$ can be controlled independently with respect to indices $s$, $i$, and $j$ because each coupling is created via an independent PA laser.
Setting $\nu_{s}$ to be bimodal, $\sqrt{n_{\rm ms}}\sigma_{\nu}$ for even $s$ and $-\sqrt{n_{\rm ms}}\sigma_{\nu}$ for odd $s$, and the distribution function of $g_{s,ij}$ to be Gaussian with the variance $\sigma^2=\sigma_g^2$, the coupling $J_{ij,kl}\equiv (2N)^{3/2} \sum_{s}g_{s,ij}g_{s,kl}/\nu_s$ becomes Gaussian random for sufficiently large $n_{\rm ms}$ as shown in the previous section.

It is useful to summarize the necessary conditions for this scheme to be valid in terms of the several relevant energy scales,
\begin{eqnarray}
&&\!\!\!\!\!\!\!\!  \max(t_i) \lesssim 1/\tau_{\rm exp} \ll J,
\label{eq:cond1}
\\
&&\!\!\!\!\!\!\!\!   \max(\Gamma_{\rm PA},\Gamma_{{\rm ms},s})\ll |\nu_{s}| \ll \Delta_{\rm min}, \,{\rm for}\,\,{\rm all}\,\, s,
\label{eq:cond2}
\\
&&\!\!\!\!\!\!\!\!   \Delta_{\rm max} < \Delta_{\rm MB} < \tilde{\Delta},
\label{eq:cond3}
\\
&&\!\!\!\!\!\!\!\!   |\nu_{s}| \ll |U_{s,s'}|, \,{\rm for}\,\,{\rm all}\,\, s \,\, {\rm and}\,\, s',
\label{eq:cond4}
\\
&&\!\!\!\!\!\!\!\!   |U_{s,s'}| < \Delta_{\rm min}\,\, {\rm or}\,\, \Delta_{\rm max} < |U_{s,s'}|,
\,{\rm for}\,\,{\rm all}\,\, s \,\, {\rm and}\,\, s'.
\label{eq:cond5}
\end{eqnarray}
Here, $t_i$ and $\tau_{\rm exp}$ denote the intersite hopping for atoms in the $i$-th band and the lifetime of the experimental system, which is approximately a few seconds in typical experiments of ultracold gases in optical lattices. Notice that the total number of necessary PA lasers in this scheme is $n_{\rm ms}\times N(N-1)/2$. 
In Appendix D, we discuss the feasibility of this proposed scheme by taking $^6$Li and a double-well optical lattice~\cite{sebby-06,anderlini-07} (see also Fig.~\ref{fig:dwl}),
\begin{eqnarray}
V_{\rm ol}({\bf r}) &=& V_0
\biggl[
\cos^2 \left( \frac{\pi x}{a} \right) + \sin^2 \left( \frac{\pi y}{a}\right)\biggr.
\nonumber \\
&& \left. + R \left( \cos\left( \frac{\pi x}{a} - \theta \right) + \cos\left( \frac{\pi y}{a} \right) \right)^2
\right],
\label{eq:DWL}
\end{eqnarray}
as specific choices of atomic species and lattice configuration. 

\section{Measuring observables}
\label{sec:scrambling}
\begin{figure}[tb]
\centering
\includegraphics[scale=0.6]{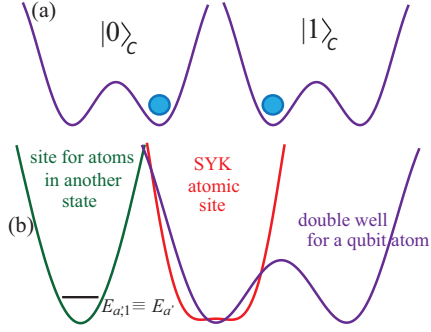}
\caption{\label{fig:otoc}
Schematic illustrations of the qubit states (a) and the configulation for measuring the OTOC functions of Eq.~(\ref{eq:OTOC}) (b).  
}
\end{figure}
\begin{figure}[tb]
\centering
\includegraphics[scale=0.5]{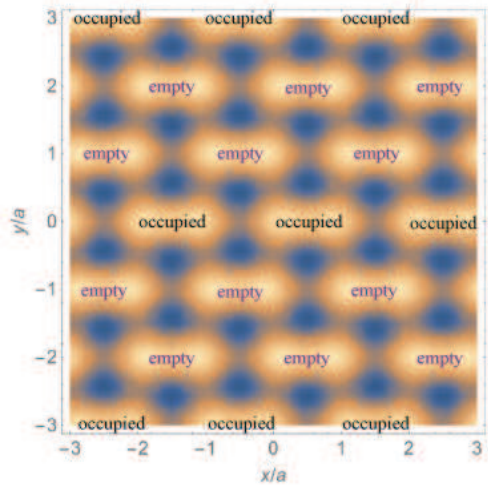}
\caption{\label{fig:qdwl}
Spatial profile of the optical lattice of Eq.~(\ref{eq:DWLq-bit}) for qubit atoms, where $V_0'<0$ and $R=0.3$. The dark and light colors indicate the high- and low-potential regions. This means that the lightest spots correspond to the sites for the qubit atoms.
}
\end{figure}
Once the SYK model is realized in optical-lattice systems, various observables can be measured. 
One of the most interesting signatures of a black hole formation is the fast scrambling
quantified by OTOC functions~\cite{Maldacena:2015waa}, with which the chaotic nature of the system can be studied quantitatively. 
Recently, Swingle {\it et al.}~have proposed a general protocol to measure the OTOC functions~\cite{swingle-16},
\begin{eqnarray}
F(t) = \langle \hat{W}^{\dagger}(t)\hat{V}^{\dagger}(0)\hat{W}(t)\hat{V}(0)\rangle.
\end{eqnarray}
We follow the protocol to explain how to measure the OTOC functions in our system in the specific case that $\hat{V}=\hat{c}_i$ and $\hat{W}=\hat{c}_j$, namely
\begin{eqnarray}
C_{i,j}(t)=\langle \hat{c}_j^{\dagger}(t)\hat{c}_i^{\dagger}(0)\hat{c}_j(t)\hat{c}_i(0) \rangle.
\label{eq:OTOC}
\end{eqnarray}
Since the SYK model is homogeneously random and has no meaningful distance, this correlation function takes only two different cases, namely the onsite case ($i=j$) and the offsite case ($i\neq j$). Hence, it is sufficient to show the cases that $i, j\in\{1,N\}$. 

The protocol requires a control qubit interacting with the probed system~\cite{swingle-16}. We assume that a double well occupied by a single atom plays the role of a control qubit and regard the state in which the atom occupies the right (left) well as the $|0\rangle_{\mathcal C}$ ($|1\rangle_{\mathcal C}$) state of the qubit as shown in Fig.~\ref{fig:otoc}(a). We also assume that the species of the qubit atoms is different from that of the SYK atoms and that the optical lattice potential for the former can be controlled independently of that for the latter. We locate the qubit double well in such a way that its left well is well overlapped with the site for the SYK atoms [see Fig.~\ref{fig:otoc}(b)]. In this situation, the qubit atom has the onsite interactions $\tilde{U}_i$ with the SYK atom in band $i$ when the qubit state is $|1\rangle_{\mathcal C}$. Specifically, supposing that the optical lattice for the SYK atoms is given by Eq.~(\ref{eq:DWL}), that for the qubit atoms may be formed by the following double-well optical lattice~\cite{sebby-06,anderlini-07},
\begin{eqnarray}
V_{\rm qb}({\bf r}) &=& V_0'
\biggl[
\cos^2 \left( \frac{\pi x}{a} \right) + \cos^2 \left( \frac{\pi y}{a}\right)
\biggr.
\nonumber \\
&& \left. + R' \left( \sin\left( \frac{\pi x}{a} \right) + \cos\left( \frac{\pi y}{a} \right) \right)^2
\right],
\label{eq:DWLq-bit}
\end{eqnarray}
whose spatial distribution for $V_0'<0$ and $R'=0.3$ is depicted in Fig.~\ref{fig:qdwl}.

The protocol to measure $C_{i,j}(t)$ is summarized as follows:
\begin{enumerate}
\renewcommand{\labelenumi}{(\roman{enumi})}
\item Prepare $(|0\rangle_{\mathcal C} + |1\rangle_{\mathcal C})/\sqrt{2}$,
\item $\hat{I}_{\mathcal S}\otimes |0\rangle\langle0|_{\mathcal C} + \hat{c}_i\otimes |1\rangle\langle1|_{\mathcal C}$,
\item $e^{-i\hat{H}t} \otimes \hat{I}_{\mathcal C}$,
\item $\hat{c}_{j}\otimes \hat{I}_{\mathcal C}$,
\item $e^{i\hat{H}t}\otimes \hat{I}_{\mathcal C}$,
\item $\hat{c}_i \otimes |0\rangle\langle0|_{\mathcal C} + \hat{I}_{\mathcal S}\otimes |1\rangle\langle1|_{\mathcal C}$,
\item Measure $\hat{X}_{\mathcal C}$ or $\hat{Y}_{\mathcal C}$,
\end{enumerate}
where $\hat{I}$ denotes the identity matrix. $\hat{X}_{\mathcal C}$ and $\hat{Y}_{\mathcal C}$ denote the $x$ and $y$ components of the Pauli matrices for the control qubit. Taking the offsite case that $i=1$ and $j=N$, let us elaborate this protocol item by item. The onsite case can be treated in a very similar way. First, since $(|0\rangle_{\mathcal C} + |1\rangle_{\mathcal C})/\sqrt{2}$ is the ground state of an atom in the symmetric double well for $\tilde{U}_i=0$, it can be straightforwardly prepared by turning off $\tilde{U}_i$ with use of the Feshbach resonance. 

In process (ii), we need to annihilate an atom at site $i$ when the qubit state is $|1\rangle_{\mathcal C}$. For this purpose, we prepare a lattice site for 
an atom in another state $a'$, which is neighboring to the SYK site as shown in Fig.~\ref{fig:otoc}(b). This state $a'$ may be a different hyperfine state or an electronically excited state as long as the linewidth of the state is sufficiently small. At $\tau=0$, where $\tau$ denotes the present time during the protocol, the interaction between the qubit atom and the SYK atom 
is set to be attractive, i.e., $\tilde{U}_i<0$. At the same time, we apply a $\pi$-pulse with frequency $\omega = E_{a'}-E_{a,1}-\tilde{U}_1$, which resonantly couples the $E_{a,1}$ and $E_{a'}$ states 
if the qubit state is $|1\rangle_{\mathcal C}$. Here $E_{a'}$ denotes the energy of the lowest band of the atoms in another state. Hence, the application of the $\pi$-pulse leads to the operation of $\hat{c}_1$ for the $|1\rangle_{\mathcal C}$ state and that of the identity matrix for the $|0\rangle_{\mathcal C}$ state, thus creating process (ii).

We next turn off the atomic interaction and perform the unitary time evolution of the system until $\tau = t$. This corresponds to process (iii). At $\tau = t$, 
we apply a $\pi$-pulse with frequency $\omega = E_{a'} - E_{a,N}$, which resonantly couples the $E_{a,N}$ and $E_{a'}$ states. This corresponds to the operation of $\hat{c}_N$, namely process (iv).

Process (v) requires the sign of the Hamiltonian to be inverted. 
Such a manipulation can be made for our SYK model simply by inverting the sign of the detuning for all the PA lasers. We perform the unitary time evolution of the inverted Hamiltonian until $\tau =2t$. At $\tau = 2t$, we set the atomic interaction to be repulsive ($\tilde{U}_{i}>0$) and apply a $\pi$-pulse with frequency $\omega = E_{a'} - E_{a,1}$, which leads to the operation of $\hat{c}_1$ for the $|0\rangle_{\mathcal C}$ state and no operation for $|1\rangle_{\mathcal C}$. This is nothing but process (vi).

In order to obtain $\langle\hat{X}_{\mathcal C}\rangle$, we need to measure the population of the bonding state
in the qubit double well. Such a measurement can be made by means of the band-mapping techniques used in Ref.~\cite{anderlini-07}.
On the other hand, in order to obtain $\langle \hat{Y}_{\mathcal C}\rangle$, we need to measure the current between the two wells, which is possible using the optical lattice microscope techniques~\cite{gemelke-09, bakr-09, sherson-10, hung-10, fukuhara-13}. Thus, process (vii) is feasible.

It has been also suggested that the degeneracy of the ground state in the SYK model can be read off from the low-temperature behavior of the single-particle Green's functions~\cite{Sachdev:2015efa},
\begin{eqnarray}
G_{i,j}(t) = \langle \hat{c}_{j}^{\dagger}(t)\hat{c}_i(0) \rangle.
\end{eqnarray}
We suggest that $G_{i,j}(t)$ can be measured in a way very similar to the one described above. Specifically, skipping processes (iv) and (v) corresponds to a protocol to measure $G_{i,i}(t)$. The offsite case ($i\neq j$) is also possible simply by replacing $\hat{c}_{i}$ operation in process (vi) with $\hat{c}_{j}$.

\section{Bottlenecks and possible solutions}
While we have shown in the previous sections that the SYK model can be realized in principle by using ultracold fermionic atoms in optical lattices coupled via PA lasers, there remain difficulties in practice, which have to be resolved in order to realize actual experiments. In this section, we describe such difficulties and discuss possible solutions to them.

First, the severest bottleneck is the number of necessary PA lasers, $n_{\rm ms}\times N(N-1)/2$. For instance, when $n_{\rm ms}=36$ and $N=16$, 4320 PA lasers are required. Implementation of lasers with as many frequencies as $O(10^3)$ in a single experiment is difficult with current experimental technology. A possible way to circumvent this difficulty is as follows. The Gaussian randomness of the coupling $J_{ij,kl}$ in the SYK model, which requires the use of multiple molecular states leading to the factor of $n_{\rm ms}$ in the number of necessary PA lasers, might be needed only to make the theory analytically solvable. In other words, the modified SYK model of Eq.~(\ref{Hamiltonian_AMO}) with only one or a few molecular states may exhibit the SY state at low temperatures. Indeed, in a supersymmetric generalization of the SYK model, which has been proposed very recently, the system exhibits the SY state even though the coupling $J_{ij,kl}$ is not Gaussian random~\cite{fu-17}. In future theoretical studies, it will be important to examine the robustness of the SY state in the absence of the Gaussian randomness in more general situations without supersymmetry. If the SY state can survive at $n_{\rm ms}=O(1)$, the number of necessary PA lasers for $N\gtrsim 10$ will be reduced to $O(10^2)$. Even in this case, preparing such a number of PA lasers remains as a challenge.


The second bottleneck is that all of the multiple PA lasers have to have ultranarrow linewidth. This requirement stems from the condition (\ref{eq:cond2}), meaning that the linewidth has to be much smaller than the detuning $|\nu_s|$. As shown in Appendix D, if we choose the double-well optical lattice of Eq.~(\ref{eq:DWL}) and set $N=16$, $V_0=-60E_{\rm R}$, $R=0.59$, and $\theta=\pi/6$, then the minimum level spacing is given by $\Delta_{\rm min}=h\times 66.7\,{\rm Hz}$ such that $\Gamma_{\rm PA}\lesssim 2\pi \times 1\,{\rm Hz}$ is required. State-of-art experiments have successfully stabilized a laser with a single frequency to the extent that $\Gamma_{\rm PA}\sim 2\pi \times 0.1 \,{\rm Hz}$, aiming to its application to optical-lattice atomic clocks~\cite{nakajima-10,inaba-14,takamoto-15}. However, achieving such narrow linewidth for lasers with multiple frequencies is challenging for current experiments. An alternative route to circumvent this difficulty may be to design a new configuration of optical lattice optimized for enlarging $\Delta_{\rm min}$ significantly compared to the case of the double-well optical lattice.

Third, it is unclear which molecular states are suited for our purpose because some information regarding molecular properties is currently unknown. Specifically, while it is more desirable to have stronger coupling between atomic and molecular states, i.e., larger Franck-Condon factor, we do not know which molecular states have relatively stronger coupling. Information regarding linewidths of electronically ground-state molecular states is insufficient as well. Moreover, in order to satisfy the conditions (\ref{eq:cond4}) and (\ref{eq:cond5}), we need to confirm that $|U_{s,s'}|$ is not too small by accident but currently we do not know the values of the s-wave scattering lengths determining the interaction between two molecules. These unknown properties can be revealed in a step-by-step manner with current experimental technology while it requires a lot of efforts.

Fourth, while we assumed that the optical-lattice depth for molecules can be controlled independently of that for atoms, such a situation has not been realized in experiments thus far. However, assuming that a PA molecule consists of two electronically ground-state atoms with different hyperfine states, at least one of the two atoms forming the molecule has a hyperfine state different from atoms in the SYK system. An optical lattice whose depth can be controlled independently of two hyperfine states has been already realized in experiments~\cite{mckay-10,gadway-10}. The optical lattice of this type should also allow for independent control of the lattice depths for the atoms and molecules.

Finally, as for the measurement scheme of the time-dependent correlation functions, the preparation of qubit atoms interacting with the probed system has never been realized thus far, while some other theoretical works recently proposed similar measurement schemes using control qubits~\cite{swingle-16,mansell-14, mitchison-16}.
Specifically, although ultracold two-species mixtures have been created in many laboratries, developing optical-lattice microscope techniques for such mixtures still stands as an experimental challenge. Nevertheless, given the fact that optical-lattice microscope techniques have been rapidly developed in recent experiments for several different atomic species, including $^{87}{\rm Rb}$~\cite{bakr-09, sherson-10}, $^{174}{\rm Yb}$~\cite{miranda-15,yamamoto-16}, $^{6}{\rm Li}$~\cite{parsons-15}, and $^{40}{\rm K}$~\cite{cheuk-15,haller-15,omran-15}, it is expected that future experiments will be able to overcome this challenge.

\section{Conclusion}
We have suggested that ultracold gases in optical lattices can be applied to experimental studies of quantum gravity under the assumption of the holographic principle. As a specific example, we have proposed
that creating the SYK model, whose low-temperature state has been conjectured to be dual to AdS$_2$ black holes~\cite{Sachdev:2015efa}, is in principle possible. We have shown how to measure the OTOC functions and the single-particle Green's function, which characterize the properties of the black hole, with use of a control qubit consisting of an atom in a double well. Moreover, we have discussed practical difficulties in realizing our proposal with current experimental technology, and how they might be circumvented.
We emphasize that while our proposal to realize the SYK model in experiment is incomplete in a practical sense because of the remaining difficulties, it has made a first step towards the experimental realization of the SYK model. In this sense, the present work has significantly advanced our original idea that quantum gravity can be studied in optical-lattice systems loaded with ultracold gases with the help of the holographic principle.

We chose our specific example because it looked the simplest among the currently available models with holography. 
However, the SYK model is still rather complicated in the sense that it has an unnatural two-body hopping that has to be Gaussian random. Hence, it will be useful to explore its simplified variants or other quantum gauge models with holography that can be created more easily in optical-lattice experiments. 
We finally note that while the Hawking radiation is one of the most important issues regarding quantum black holes, whether the Hawking radiation can be seen in the SYK model is not clear at this stage. Answering to this question will be an imperative task for future theoretical studies.
At very least, it should be possible to study a variant of the information puzzle, associated to the
way that the information is encoded in a black hole~\cite{Maldacena:2001kr}.

\section*{Acknowledgment}

The authors thank S.~Aoki, G.~Gur-Ari, S.~Nakajima, M.~Sheleier-Smith, S.~Shenker, B.~Swingle, and Y.~Takahashi for discussions. Discussions during the YITP workshop (YITP-W-16-01) on ``Quantum Information in String Theory and Many-body Systems" were useful to complete this work. The authors acknowledge KAKENHI grants from JSPS: Grants No.~JP15H05855 (I.D. and M.T.), No.~JP25220711 (I.D.), No.~JP25287046 (M.H.), and No.~JP26870284 (M.T.). I.D. was supported by research grant from CREST, JST No. JPMJCR1673.




\begin{thebibliography}{9}
\bibitem{footnote2}
This Hamiltonian is the one with complex fermion \cite{Sachdev:2015efa}, 
which is slightly different from the one which uses real fermions \cite{Kitaev_talk}.   
They are equivalent at the large-$N$ limit with fixed $J$, after the disorder average is taken.
\end{thebibliography}
%

\appendix
\vspace{10mm}
$\!\!\!\!\!\!\!\!${\Large Appendices:}

\section{Comparison of the SYK model with its variants}
In this appendix, we compare the three versions of Sachdev-Ye-Kitaev (SYK) model, namely the original (complex) one, the real one, and the modified one (11) to demonstrate that the last one is a quantitative approximation of the original one even at finite $N$ and $n_{\rm ms}$. We set the reduced Planck's constant and the Boltzmann constant to be $\hbar = k_{\rm B} = 1$

\begin{figure}[tbp]
\centering
\includegraphics[scale=0.8]{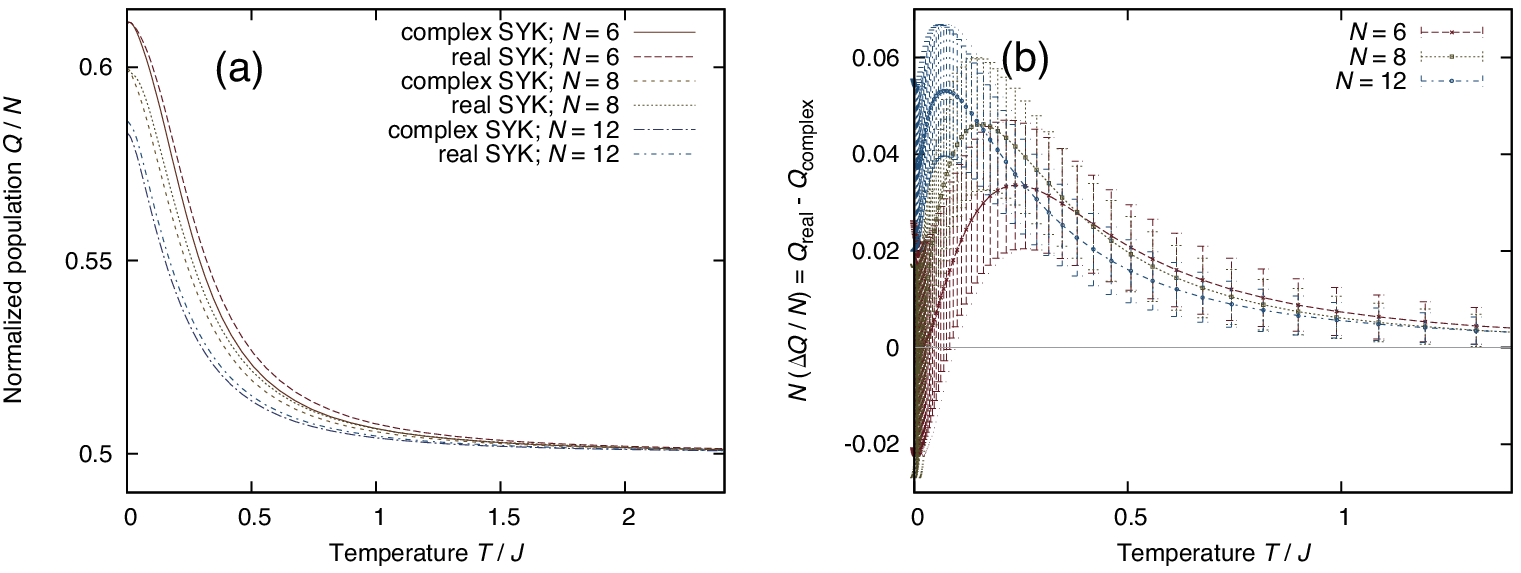}
   \caption{(a) Comparison of the $(T/J)$-dependence of $\overline{\langle Q\rangle}/N$ between the the original (complex) and real SYK models for different $N$.
	(b) The difference of $\overline{\langle Q\rangle}$ between the real and complex SYK models, which is $N$ times the difference between the values of $\overline{\langle Q\rangle}/N$.
   The chemical potential is set to zero, i.e., $\mu=0$.
	$10^3$ samples are taken for $N=6, 8$, and $12$.
   }
\label{fig:thermo_N_charge}
\end{figure}

\begin{figure}[tbp]
\centering
\includegraphics[scale=0.8]{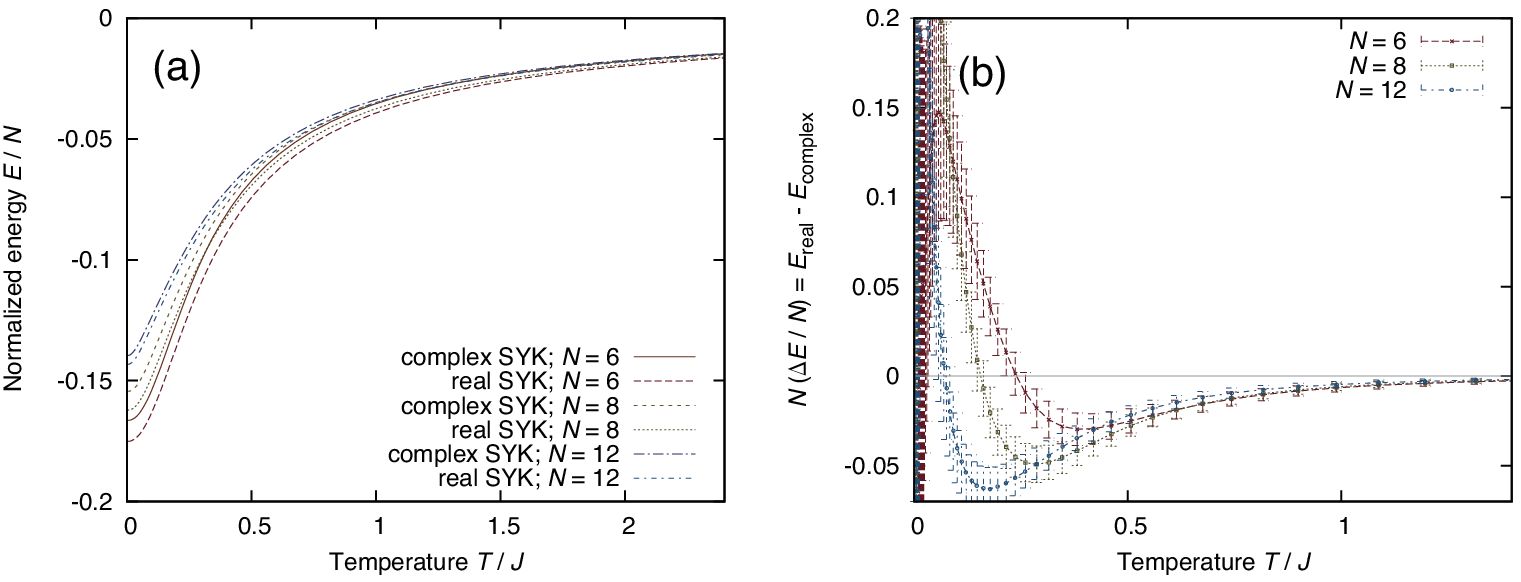}
   \caption{(a) Comparison and (b) difference of $(T/J)$-dependence of $\overline{\langle E\rangle}$ between the complex and real SYK models and different $N$.
   The chemical potential is set to zero, i.e., $\mu=0$. 
	$10^3$ samples are taken for $N=6, 8, 12$.
   }
\label{fig:thermo_real_energy}
\end{figure}

We first compare the complex and real SYK models.
In Fig.~\ref{fig:thermo_N_charge}, 
the $(T/J)$-dependence of $Q/N$ in the complex and real SYK models is shown, where $Q$ denotes the number of fermions, or the charge, defined as the eigenvalue of the number operator  $\hat{Q}=\sum_{i=1}^N\hat{n}_i=\sum_{i=1}^N\hat{c}_i^\dagger\hat{c}_i$ that commutes with all the Hamiltonians considered here.
We label the energy eigenvalues of each model Hamiltonian by the charge as  $\{E_i^{(Q)}\}_i$ and obtain
\begin{eqnarray}
\langle
Q
\rangle_{T,J}
=
\frac{\sum_Q Q\cdot Z^{(Q)}}{\mathcal{Z}},
\end{eqnarray}
in which $Z^{(Q)}$ is the canonical partition function $Z^{(Q)} = \sum_i e^{-E_i^{(Q)}/T}$ and $\mathcal{Z} = \sum_Q Z^{(Q)}$ is the grandcanonical partition function.
In Fig.~\ref{fig:thermo_real_energy}, 
the $(T/J)$-dependence of the energy in the complex and real-SYK models, normalized by dividing by $N$, are shown. 
The energy $E$ is calculated as 
\begin{eqnarray}
\langle
E
\rangle_{T,J}
=
\frac{\sum_{Q,i}
E_i^{(Q)}\cdot e^{-E_i^{(Q)}/T}}
{\mathcal{Z}}.
\end{eqnarray}
The disorder average $\overline{\langle E\rangle}$ and $\overline{\langle Q\rangle}$ are taken by using random couplings.
From the plots, we can see a clear agreement at large $N$. 
As expected from the standard $1/N$-counting, two theories give the same values of $\overline{\langle E\rangle}$ and $\overline{\langle Q\rangle}$ up to the sub-leading corrections of order $N^0$.


\begin{figure}[tbp]
\centering
     \includegraphics[scale=0.8]{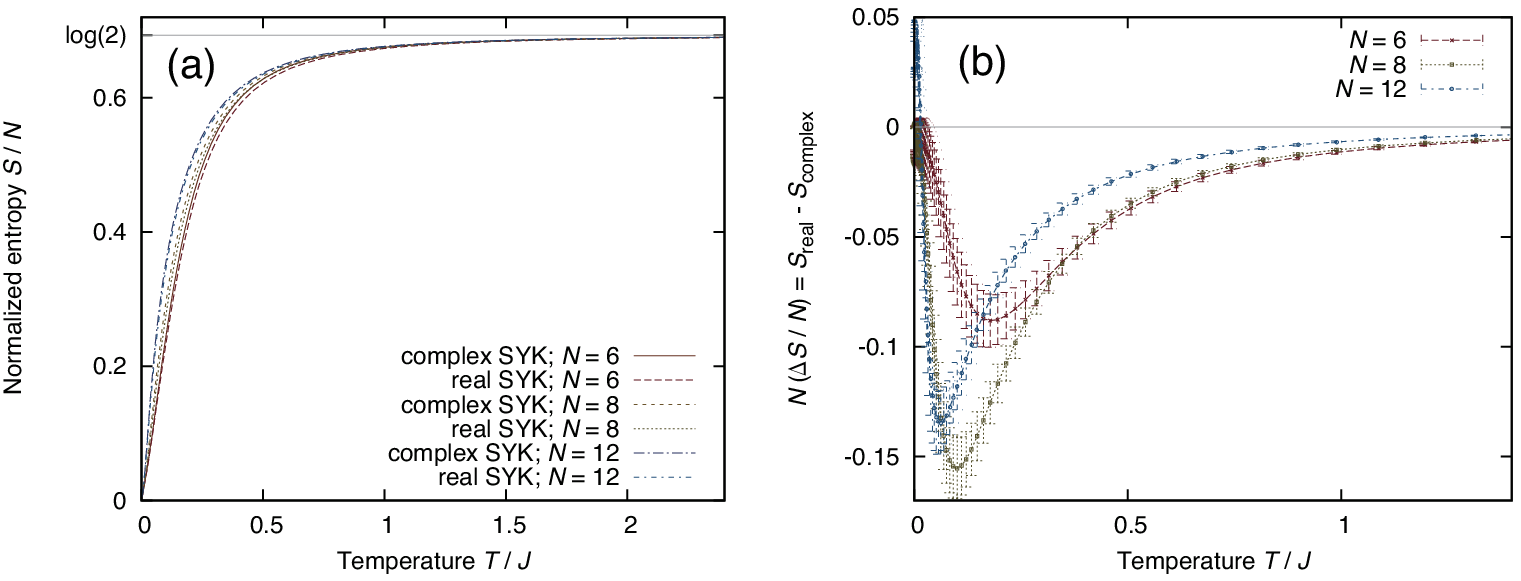}
   \caption{(a) Comparison and (b) difference of $(T/J)$-dependence of $\overline{\langle S\rangle}/N$ between the complex and real SYK models and different $N$.
   The chemical potential is set to zero, $\mu=0$. 
The entropy approaches $S = N\log 2$ as the temperature $T$ is increased, which is expected from the fact that all $2^N$ states can equally contribute in the high-temperature limit.
   }
\label{fig:thermo_real_vs_complex_entropy}
\end{figure}

We have also calculated the entropy $S$ defined by 
\begin{eqnarray}
\overline{S}=\frac{\overline{\langle E\rangle}}{T}+\overline{\log \mathcal{Z}}. 
\end{eqnarray}
Note that $\overline{\log \mathcal{Z}} < \log \overline{\mathcal{Z}}$ in general.
The result is shown in Fig.~\ref{fig:thermo_real_vs_complex_entropy}. 
For $T/J\gtrsim 1$, $S/N$ is already almost converged at $N=6$, while for smaller $T$, $S/N$ is an increasing function of $N$. Notice that the entropy of the complex SYK model at finite $N$ has been computed also in Ref.~\cite{fu-16}.

\begin{figure}[tbp]
\centering
     \includegraphics[scale=0.8]{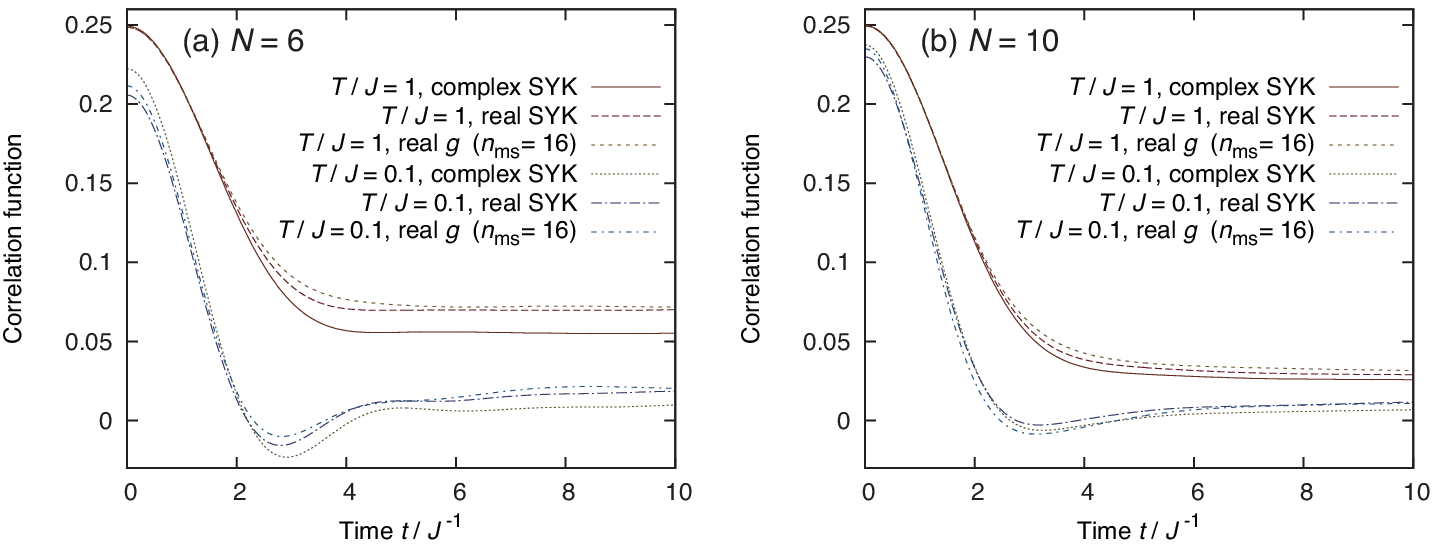}
   \caption{The real-time, same-site density-density correlation function $\sum_i \overline{\langle\hat{n}_i(t)\hat{n}_i(0)\rangle-\langle\hat{n}_i(t)\rangle\cdot\langle\hat{n}_i(0)\rangle}/N$
calculated for (a) $N=6$ using $10^3$ samples, and (b) $N=10$ using $10^2$ samples. The data for $T=1$ and $0.1$ for the complex and real SYK models are plotted together with
the data for the model (11). 
}
\label{fig:nn-correlator}
\end{figure} 

As an example of a two-point function,
in Fig.~\ref{fig:nn-correlator} we present the same-site density-density correlation function $C_\mathrm{nn}(t)$, which is defined using the number operator $\hat{n}_i = \hat{c}_i^\dag \hat{c}_i$ by
\begin{eqnarray}
C_\mathrm{nn}^{(i)}(t)
= \overline{\langle\hat{n}_i(t)\hat{n}_i(0)\rangle-\langle\hat{n}_i(t)\rangle\cdot\langle\hat{n}_i(0)\rangle}
= \overline{\langle\hat{n}_i(t)\hat{n}_i(0)\rangle}-\overline{\langle\hat{n}\rangle^2}
\equiv \overline{\langle\hat{n}_i(t)\hat{n}_i(0)\rangle_\mathrm{conn}};
\end{eqnarray}
\begin{eqnarray}
C_\mathrm{nn}(t) = \frac{1}{N}\sum_i C_\mathrm{nn}^{(i)}(t).
\end{eqnarray}
Here, the operator $\hat{\mathcal{O}}(t)$ at time $t$ is $\hat{\mathcal{O}}(t) = e^{i\hat{H}t}\hat{\mathcal{O}}e^{-i\hat{H}t}$
and the expectation values are calculated as in the cases of the charge and the energy:
\begin{eqnarray}
\langle
\cdots
\rangle = \frac{\sum_{Q,j} e^{-\beta E^{(Q)}_j}\langle \psi^{(Q)}_j \vert 
\cdots
\vert \psi_j^{(Q)} \rangle}{\mathcal{Z}}
\end{eqnarray}
with $\vert \psi_j^{(Q)} \rangle$ being the many-body wavefunction corresponding to the eigenenergy $E^{(Q)}_j$.

\begin{figure}[tbp]
\centering
     \includegraphics[scale=0.8]{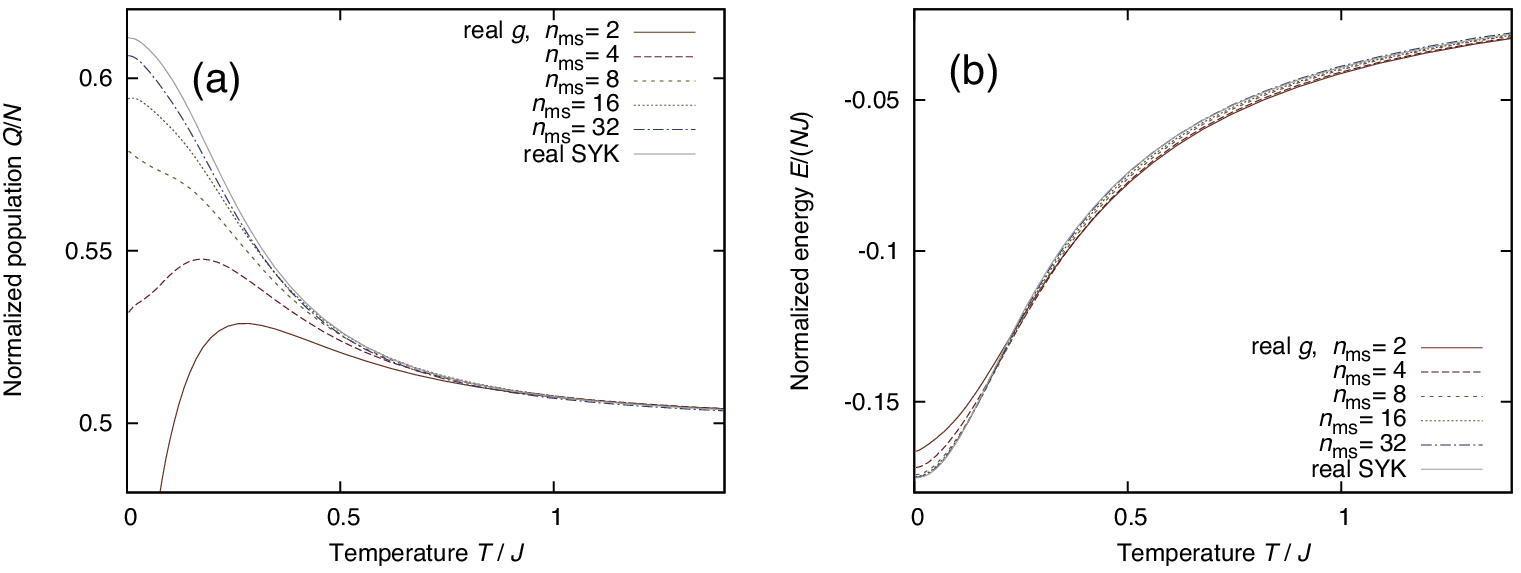}
   \caption{The $(T/J)$-dependence of (a) $\overline{\langle Q\rangle}$ and (b) $\overline{\langle E\rangle}$
	for $N = 6$ for $n_{\rm ms} = 2, 4, 8, 16, 32$. 
    The chemical potential is set to zero, $\mu=0$. 
    The results for the real SYK model are shown for comparison.
    $10^4$ samples are used and the standard error estimates (not shown) are typically on the order of the width of the lines.
   }
\label{fig:thermo_real_charge_energy}
\end{figure} 
\begin{figure}[tbp]
\centering
     \includegraphics[scale=0.8]{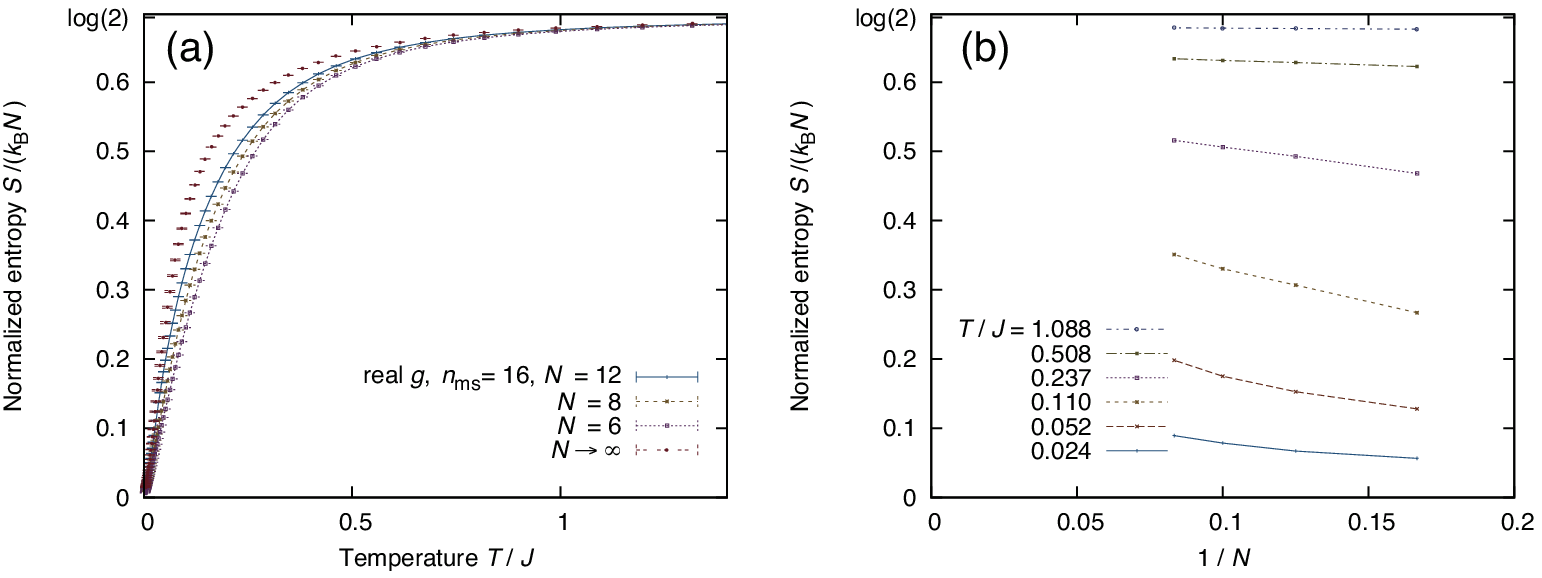}
   \caption{(a) Comparison of the $(T/J)$-dependence of $\overline{S}/(k_\mathrm{B} N)$, the modified SYK model (11) with $n_{\rm ms} = 16$, extrapolated to $N=\infty$ by a linear fit of the value against $1/N$.
The chemical potential is set to zero, $\mu=0$. 
(b) $\overline{\langle S\rangle}/(k_\mathrm{B} N)$ plotted against $1/N$ for $N=6,8,10,12$.
For higher $T$ the obtained normalized entropy is linear in $1/N$, however for $T/J\ll 1$ the curve is more convex, which suggests that the actual $N\to\infty$ limit may be significantly larger than the value plotted and may converge to a finite value as $T\to 0$.}
\label{fig:thermo_real_vs_complex_entropy_large-N}
\end{figure} 

Having corroborated the quantitative agreement between the complex and real SYK models, we next compare the modified SYK model of Eq.~(11) with the real SYK model.
As shown in Fig.~\ref{fig:thermo_real_charge_energy} for $\overline{\langle Q\rangle}$ and $\overline{\langle E\rangle}$,
we observe that the results are already similar for $n_{\rm ms}=8$ for $N=6$ and $10$.
In Fig.~\ref{fig:thermo_real_vs_complex_entropy_large-N} (a) we plot the entropy $\overline{S}$ as a function of the temperature for $N=6, 8, 12$ for $n_{\rm ms} = 16$,
along with the result of linear extrapolation to $1/N\to 0$. As in Fig.~\ref{fig:thermo_real_vs_complex_entropy_large-N} (b), for $T\gtrsim 0.1J$ the obtained entropy is almost linear in $1/N$, however for lower temperatures the dependence of $\overline{S}$ on $1/N$ is more convex, indicating that the linear fit from $N=6, 8, 12$ may be underestimating the value of $\overline{S}(N\to\infty)$ at $T\to 0$ and that $\overline{S}$ may converge to a finite value.

It is worth stressing the importance of large $n_{\rm ms}$. Even when $n_{\rm ms}=1$, the model (11) looks like the real-SYK model if we identify $J_{ij,kl}/(2N)^{3/2}$ with $g_{ij}g_{kl}/\nu$, where $g_{ij}\equiv g_{1,ij}$ and $\nu\equiv \nu_1$. However, the distribution of the latter is not Gaussian in general for a given distribution of $\{g_{ij}\}$ and even worse, i.e., the randomness is not strong enough; for example, when $g_{12}g_{12}/\nu$ and $g_{34}g_{34}/\nu$ are large, $g_{12}g_{34}/\nu$ is also large. Note also that $g_{ij}g_{ij}$ is always positive. 

\section{Properties of $J_{ij,kl}=(2N)^{3/2}\sum_{s=1}^{n_{\rm ms}}g_{s,ij}g_{s,kl}/\nu_s$}\label{app:J=gg}
\hspace{0.51cm}
In this appendix, we explain basic properties of $J_{ij,kl}\equiv(2N)^{3/2}\sum_{s=1}^{n_{\rm ms}}\frac{g_{s,ij}g_{s,kl}}{\nu_s}$. 
We take $\nu_s=+\sqrt{n_{\rm ms}}\sigma_{\nu}$ for even $s$ and $\nu_s=-\sqrt{n_{\rm ms}}\sigma_{\nu}$ for odd $s$, 
and the Gaussian weight of $g_{s,ij}$ is chosen to be $\frac{e^{-g_{s,ij}^2/(2\sigma_\mathrm{g}^2)}}{\sqrt{2\pi}\sigma_\mathrm{g}}$,
with $\sigma_\mathrm{g}^2 / \sigma_{\nu} = (2N)^{-3/2}J$.
It will turn out that this corresponds to the Gaussian random coupling $J_{ij,kl}$ needed for the real-SYK model, 
with $J=1$. Generic values of $J$ can be realized by rescaling $g_{s,ij}$ and/or $\nu_s$.  

\begin{figure}[tbp]
\centering
     \includegraphics[scale=0.8]{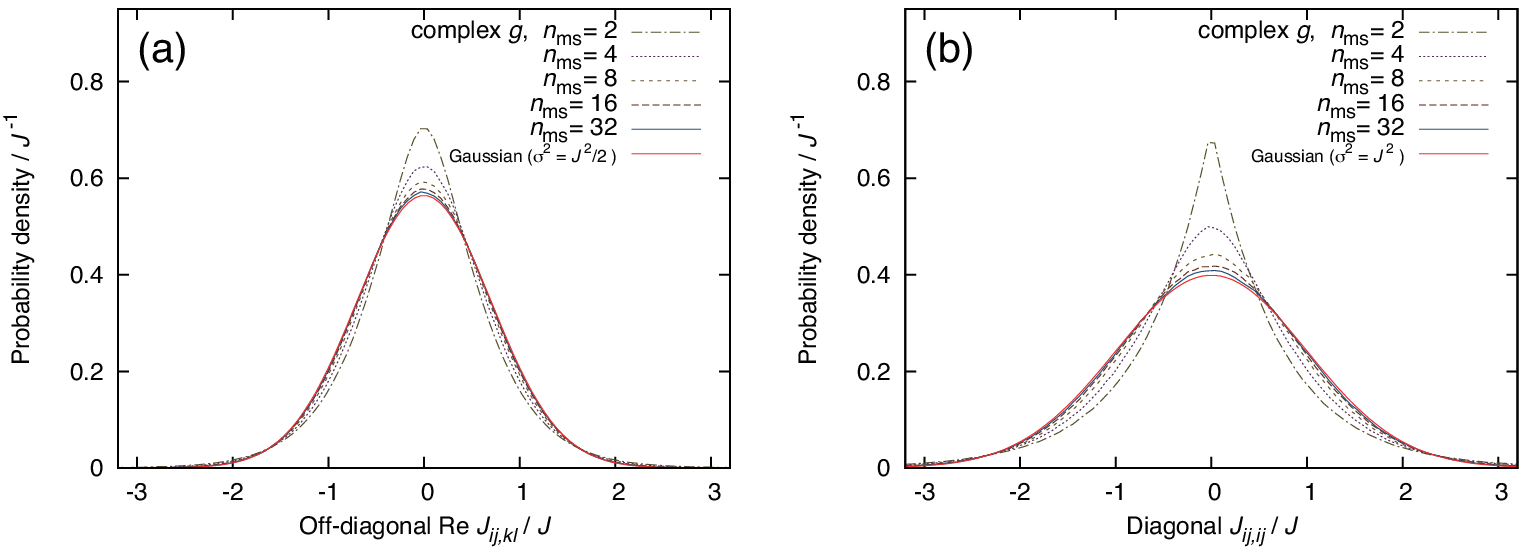}
   \caption{(a): Distribution of $J_{ij,kl}=\frac{(2N)^{3/2}}{\sqrt{n_{\rm ms}}J}\left(\sum_{s:{\rm even}}g_{s,ij}g_{s,kl}^{\ast}-\sum_{s:{\rm odd}}g_{s,ij}g_{s,kl}^{\ast}\right)$ with only the off-diagonal components (i.e. $(i,j)\neq (k,l),(l,k)$); 
(b): Distribution of the real $J_{ij,ij}=\frac{(2N)^{3/2}}{\sqrt{n_{\rm ms}}J}\left(\sum_{s:{\rm even}}\vert g_{s,ij}\vert^2-\sum_{s:{\rm odd}}\vert g_{s,ij}\vert^2\right)$.
   The weights of ${\rm Re} g_{s,ij}$ and ${\rm Im} g_{s,ij}$ are Gaussian with variance $\sigma_\mathrm{g}^2 = (2N)^{-3/2} J^2/2$, $\frac{e^{-|{\rm Re}({\rm Im})g_{s,ij}|^2/(\sigma_\mathrm{g}^2)}}{\sqrt{\pi}\sigma_\mathrm{g}}$. 
   The distribution of ${\rm Re} J_{ij,kl}$ converges to $\frac{e^{-({\rm Re} J_{ij,kl})^2/(J^2)}}{\sqrt{\pi}J}$, which is shown in (a) as ``Gaussian ($\sigma^2=J^2/2$)''.
   The distribution of $J_{ij,ij}$ converges to the standard normal distribution (for $J=1$), $\frac{e^{-J{ij,ij}^2/(2J^2)}}{\sqrt{2\pi}J}$, as shown in (b).
The numbers of samples taken are $10^4$ (a) and $10^5$ (b), respectively. 
   }
\label{fig:dist_gg_N10}
\end{figure} 

Firstly let us show that the distribution of $x\equiv J_{ij,kl}$ converges to $e^{-x^2}/\sqrt{\pi}$ for $(i,j)\neq (k,l)$ and 
 $e^{-x^2/2}/\sqrt{2\pi}$ for $(i,j)=(k,l)$.
Then, we should show that, when real numbers $x_s$ and $y_s$ are distributed with the weight $\frac{e^{-x_s^2/2}}{\sqrt{2\pi}}$
and $\frac{e^{-y_s^2/2}}{\sqrt{2\pi}}$, (1) $\frac{1}{\sqrt{n_{\rm ms}}}\sum_{s=1}^{n_{\rm ms}}x_sy_s$, and (2) $\frac{1}{\sqrt{2n_{\rm ms}}}\sum_{s=1}^{n_{\rm ms}}(x_s^2-y_s^2)$
converge to Gaussian distribution with width $1$ and $\sqrt{2}$. 
The statements (1) and (2) are actually equivalent; indeed, by using $X_s\equiv\frac{x_s+y_s}{\sqrt{2}}$ and $Y_s\equiv\frac{x_s-y_s}{\sqrt{2}}$, 
we can rewrite the former as $x_sy_s=(X_s^2-Y_s^2)/2$ with the same weight, $e^{-(x_s^2+y_s^2)/2}=e^{-(X_s^2+Y_s^2)/2}$. 
Hence we consider only the former. Because the sum with respect to $s$ can be regarded as a random walk, 
the distribution should be Gaussian. 
Then, in order to determine the width, we only have to calculate the average of $\left(\frac{1}{\sqrt{n_{\rm ms}}}\sum_{s=1}^{n_{\rm ms}}x_sy_s\right)^2$. 
It can be evaluated as 
\begin{eqnarray}
\left\langle\left(\frac{1}{\sqrt{n_{\rm ms}}}\sum_{s=1}^{n_{\rm ms}}x_sy_s\right)^2\right\rangle
=
\left\langle\frac{1}{n_{\rm ms}}\sum_{s=1}^{n_{\rm ms}}x_s^2y_s^2\right\rangle
=
\frac{1}{\pi}\int x^2y^2e^{-(x^2+y^2)/2}dxdy
=
1, 
\end{eqnarray}
which means the width is $1$.

We can also show 
$\overline{J_{ij,kl}J_{pq,rs}}\propto(\delta_{ip}\delta_{jq}-\delta_{iq}\delta_{jp})(\delta_{kr}\delta_{ls}-\delta_{ks}\delta_{lr})
+(ij\leftrightarrow kl)$. 
 Let us note that we only have to show 
$\overline{J_{ij,kl}J_{pq,rs}}\propto\delta_{ip}\delta_{jq}\delta_{kr}\delta_{ls} +(ij\leftrightarrow kl)$ for $i<j$, $k<l$, $p<q$ and $r<s$. 
It is equivalent to show that $\overline{(\sum_s  g^{(s)}_I g^{(s)}_J/\nu_s)(\sum_{s'} g^{(s')}_P g^{(s')}_Q/\nu_{s'})}=0$ 
unless $I=P, J=Q$ or $I=Q, J=P$,  
where indices $I,J,P,Q$ represents $(i,j),(k,l),(p,q)$ and $(r,s)$. 
With this notation, 
\begin{eqnarray}
\overline{(\sum_s g^{(s)}_I g^{(s)}_J/\nu_{s})(\sum_{s'} g^{(s')}_P g^{(s')}_Q/\nu_{s'})}
=
\sum_{s,s'}\overline{g^{(s)}_I g^{(s)}_J g^{(s')}_P g^{(s')}_Q/(\nu_{s}\nu_{s'})}. 
\end{eqnarray}
If $I\neq J$ or $P\neq Q$, we can rewrite it as 
\begin{eqnarray}
\sum_{s,s'}\overline{g^{(s)}_I g^{(s)}_J g^{(s')}_P g^{(s')}_Q/(\nu_{s}\nu_{s'})} 
=
\frac{1}{n_{\rm ms}}\sum_{s}\overline{g^{(s)}_I g^{(s)}_J g^{(s)}_P g^{(s)}_Q}, 
\end{eqnarray}
where we used the invariance of the Gaussian weight w.r.t. a flip of sign of any of $g^{(s)}_I, g^{(s)}_J g^{(s')}_P, g^{(s')}_Q$ and 
$\nu_s^2=n_{\rm ms}$. 
Then, again due to the invariance of the weight w.r.t. a flip of sign of any of $g^{(s)}$, unless $I=P, J=Q$ or $I=Q, J=P$ the average vanishes. 
When $I=J\neq P=Q$, 
\begin{eqnarray}
\overline{(\sum_s g^{(s)}_I g^{(s)}_J/\nu_{s})(\sum_{s'} g^{(s')}_P g^{(s')}_Q/\nu_{s'})}
=
\overline{(\sum_s g^{(s)}_I g^{(s)}_J/\nu_{s})}\cdot\overline{(\sum_{s'} g^{(s')}_P g^{(s')}_Q/\nu_{s'})} 
=
0\cdot 0 = 0. 
\nonumber\\
\end{eqnarray}

It is a bit tricky to show $\overline{JJJ}=0$; actually it holds when $n_{\rm ms}$ is infinity. 
For simplicity, let us suppose $I\neq J$, $P\neq Q$, $V\neq W$. 
Then, 
\begin{eqnarray}
\lefteqn{\overline{(\sum_s g^{(s)}_I g^{(s)}_J/\nu_{s})(\sum_{s'} g^{(s')}_P g^{(s')}_Q/\nu_{s'})(\sum_{s''} g^{(s'')}_V g^{(s'')}_W/\nu_{s''})}
}\nonumber\\
&=&
\sum_{s}\overline{g^{(s)}_I g^{(s)}_J g^{(s)}_P g^{(s)}_Qg^{(s)}_V g_W^{(s)}}/\nu_{s}^3 
\nonumber\\
&=& 
O(1/\sqrt{n_{\rm ms}})\to 0\ (n_{\rm ms}\to\infty). 
\end{eqnarray}
For the same reason, we have 
\begin{eqnarray}
\overline{JJJJ}
=
\overline{JJ}\cdot \overline{JJ}
+
O(1/n_{\rm ms}), 
\end{eqnarray}
and so on. 

We further note that, if we can introduce complex $g_{s,ij}$, we may identify $(2N)^{3/2} \sum_s \frac{g_{s,ij}g_{s,kl}^\ast}{\nu_s}$ with $J_{ij,kl}$,
with both the distributions of $J_{ij,kl}$ (see Fig.~\ref{fig:dist_gg_N10}) and other quantities discussed in the main text
quickly approaching the distributions for the complex SYK model as $n_{\rm ms}$ is increased.

\section{Derivation of the effective Hamiltonian}
In this appendix, on the basis of the degenerate perturbation theory we present a detailed derivation of the effective Hamiltonian of Eq.~(11) from the coupled atom-molecule model of Eq.~(10)  and discuss parameter regions in which the effective Hamiltonian is valid.  

Let us write the Hamiltonian in the following form,
\begin{eqnarray}
\hat{H}_m = \hat{H}_0 + \hat{V},
\end{eqnarray}
where the non-perturbative part $\hat{H}_0$ and perturbative part $\hat{V}$ are given by 
\begin{eqnarray}
\hat{H}_0 &=& \sum_{s=1}^{n_{\rm ms}} \nu_s \hat{m}_s^{\dagger} \hat{m}_s +
\sum_{s,s'} \frac{U_{ss'}}{2}\hat{m}_s^{\dagger}\hat{m}_{s'}^{\dagger} \hat{m}_{s'} \hat{m}_s,
\label{eq:H0}
\\
\hat{V} &=& \sum_{s=1}^{n_{\rm ms}} \sum_{i,j}g_{s,ij}
\left(
\hat{m}_s^{\dagger} \hat{c}_i\hat{c}_j - \hat{m}_s \hat{c}_i^{\dagger} \hat{c}_j^{\dagger}
\right).
\label{eq:Vpert}
\end{eqnarray}
We see from Eq.~(\ref{eq:H0}) that the non-perturbative energy depends only on the number of particles in each molecular states. This means that all different atomic configurations with no molecule are degenerate in $\hat{H}_0$. The non-perturbative energy of these degenerate states is given by $E_0 = 0$. We define the Hilbert subspace spanned by all these degenerate states with no molecule as $\mathcal{D}$. We note that states with a molecule or two molecules appear as virtual states in the second- or fourth-order perturbation.

In order to derive the effective Hamiltonian, we perform the Schrieffer-Wolff transformation~\cite{schrieffer-66},
\begin{eqnarray}
\hat{H}_{\rm eff} = \hat{P}_0 e^{\hat{S}} \hat{H}_m e^{-\hat{S}}  \hat{P}_0,
\end{eqnarray}
where $\hat{P}_0$ is the projection operator on $\mathcal{D}$.
To determine the transformation matrix $\hat{S}$ we require that in $e^{\hat{S}} \hat{H}_m e^{-\hat{S}}$ all the matrix elements connecting states in $\mathcal{D}$ with those outside of $\mathcal{D}$ are zero, i.e., that $e^{\hat{S}} \hat{H}_m e^{-\hat{S}}$ is block-diagonal~\cite{bravyi-11}. 
While our main purpose is to derive Eq.~(11) of the main text, which corresponds to the effective Hamiltonian up to the second-order perturbation, we here describe the terms up to the fourth order,
\begin{eqnarray}
\hat{H}_{\rm eff} \simeq \hat{H}_{\rm eff}^{(2)} + \hat{H}_{\rm eff}^{(4)},
\end{eqnarray}
in order to discuss the validity condition of the second-order approximation. Notice that the odd order terms do not exist in the effective Hamiltonian because in $\hat{V}$ all the matrix elements connecting two states with the same number of molecules are zero.

The second- and fourth-order terms can be formally written as
\begin{eqnarray}
\hat{H}_{\rm eff}^{(2)} &=& \hat{P}_0 \hat{V}\hat{\Lambda}\hat{V}\hat{P}_0,
\label{eq:H2nd}
\\
\hat{H}_{\rm eff}^{(4)} &=& \hat{P}_0 \hat{V}\hat{\Lambda}\hat{V}\hat{\Lambda}\hat{V}\hat{\Lambda}\hat{V}\hat{P}_0 
+ \frac{1}{2}
\left( 
\hat{P}_0 \hat{V}\hat{\Lambda}^2\hat{V}\hat{P}_0\hat{V}\hat{\Lambda}\hat{V}\hat{P}_0 
+\hat{P}_0 \hat{V}\hat{\Lambda}\hat{V}\hat{P}_0\hat{V}\hat{\Lambda}^2\hat{V}\hat{P}_0
\right).
\label{eq:H4th}
\end{eqnarray}
where
\begin{eqnarray}
\hat{\Lambda} = \frac{1-\hat{P}_0}{E_0 - \hat{H}_0}.
\end{eqnarray}
Substituting Eqs.~(\ref{eq:H0}) and (\ref{eq:Vpert}) into Eqs.~(\ref{eq:H2nd}) and (\ref{eq:H4th}), we obtain
\begin{eqnarray}
\hat{H}_{\rm eff}^{(2)} &=& \sum_{ijkl}\mathcal{K}_{ij,kl}
\hat{c}_i^{\dagger}\hat{c}_j^{\dagger} \hat{c}_k \hat{c}_l,
\label{eq:H2}
\\
\hat{H}_{\rm eff}^{(4)} &=& - \sum_{ii'jj'kk'll'} \tilde{\mathcal{L}}_{ii',jj',kk',ll'}
\hat{c}_i^{\dagger}\hat{c}_{i'}^{\dagger}\hat{c}_j^{\dagger}\hat{c}_{j'}^{\dagger}
\hat{c}_k\hat{c}_{k'}\hat{c}_l\hat{c}_{l'}
\nonumber \\
&& - \sum_{ii'jj'kk'll'}\mathcal{L}_{ii',jj',kk',ll'}
\hat{c}_i^{\dagger}\hat{c}_{i'}^{\dagger}\hat{c}_j\hat{c}_{j'}
\hat{c}_k^{\dagger}\hat{c}_{k'}^{\dagger}\hat{c}_l\hat{c}_{l'}
\label{eq:H4}
\end{eqnarray}
where
\begin{eqnarray}
\mathcal{K}_{ij,kl}&=&\sum_{s}\frac{g_{s,ij}g_{s,kl}}{\nu_s}, 
\label{eq:K2}
\\
\tilde{\mathcal{L}}_{ii',jj',kk',ll'} &=& \sum_{s_2s_3} \sum_{s_1 = s_2, s_3} 
\frac{g_{s_1,ii'}g_{s_2,jj'}g_{s_2,kk'}g_{s_3,ll'}}{\nu_{s_1}(\nu_{s_2}+\nu_{s_3}+U_{s_1s_2})\nu_{s_3}}, 
\\
\mathcal{L}_{ii',jj',kk',ll'}&=&\sum_{ss'} \frac{1}{2} g_{s,ii'}g_{s,jj'}g_{s',kk'}g_{s',ll'}
\left(
\frac{1}{\nu_s^2\nu_{s'}} + \frac{1}{\nu_s \nu_{s'}^2}
\right).
\label{eq:L4}
\end{eqnarray}
It is obvious that $\hat{H}_{\rm eff}^{(2)}$ is equivalent to Eq.~(11) of the main text. Since we have assumed that $|\nu_{s}| \ll |U_{ss'}|$, the first term in the right-hand side of Eq.~(\ref{eq:H4}) is much smaller than the second term. Hence, we neglect the first term and compare $\hat{H}_{\rm eff}^{(2)}$ with the second term in $\hat{H}_{\rm eff}^{(4)}$ in the following discussions. 

We recall that $g_{s,ij}$ is assumed to be Gaussian random with the standard deviation $\sigma = \sigma_g$ [see Eq.~(21) of the main text] and that $\nu_{s}$ is assumed to be $(-1)^s \sqrt{n_{\rm ms}}\sigma_{\nu}$. Combining these assumption with Eq.~(\ref{eq:K2}), we obtain
\begin{eqnarray}
\overline{\mathcal{K}_{ij,kl}}=0, \, 
\overline{\mathcal{K}_{ij,kl}^2} = \frac{\sigma_g^4}{\sigma_{\nu}^2},
\end{eqnarray}
for $\{i,j\}\neq \{k,l\}$.
Moreover, when $\{i,i'\}$, $\{j,j'\}$, $\{k,k'\}$, and $\{l,l'\}$ are not equal to one another, we obtain
\begin{eqnarray}
\overline{\mathcal{L}_{ii',jj',kk',ll'}} = 0, \,
\overline{\mathcal{L}_{ii',jj',kk',ll'}^2} = \frac{\sigma_g^8}{2n_{\rm ms}\sigma_{\nu}^6}.
\end{eqnarray}
The scale of the eigenenergies of $\hat{H}_{\rm eff}^{(2)}$ is set by $\sqrt{4N^3\overline{\mathcal{K}_{ij,kl}^2}/3!}$ while that of $\hat{H}_{\rm eff}^{(4)}$ is set by $\sqrt{8N^7\overline{\mathcal{L}_{ii',jj',kk',ll'}^2}/7!}$~\cite{maldacena-16}. In order for the second-order approximation to be valid, the former must be much larger than the latter. This condition implies that
\begin{eqnarray}
\frac{\sqrt{{}_7{\rm P}_4\times n_{\rm ms}} }{N^2} \gg \frac{\sigma_g^2}{\sigma_{\nu}^2}.
\label{eq:last-cond}
\end{eqnarray}
In Appendix D, we show that the condition of Eq.~(\ref{eq:last-cond}) can be safely satisfied in a realistic situation.

\vspace{5mm}
\section{An example: double-well optical lattice}
\begin{figure}[tb]
\centering
\includegraphics[scale=0.45]{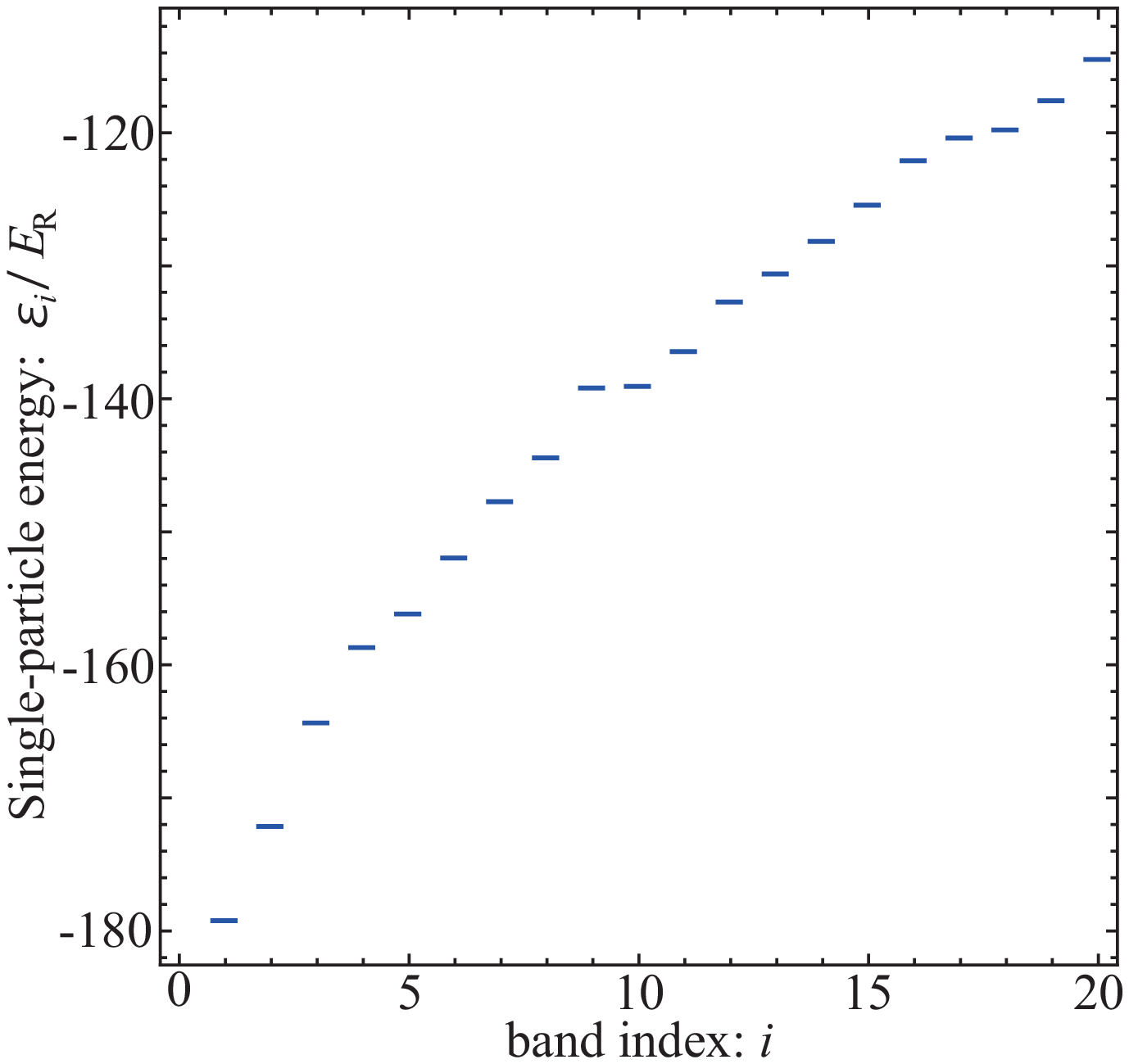}
\caption{\label{fig:Esp}
Eigenenergies of the Schr\"odinger equation for a single atom in the optical-lattice potential of Eq.~(\ref{eq:DWL}) at zero quasi-momentum, where $V_0=-60\,E_{\rm R}$, $R=0.59$, and $\theta = \pi/6$. From this energy spectrum, one can evaluate that $\Delta_{\rm min}=0.00228 \,E_{\rm R}$ and $\Delta_{\rm max}=104 \,E_{\rm R}$ for $N=16$.
}
\end{figure}

In this appendix, in order to discuss the feasibility of our scheme for creating the SYK model, we consider the following optical lattice,
\begin{eqnarray}
V_{\rm ol}({\bf r}) = V_0
\left[
\cos^2 \left( \frac{\pi x}{a} \right) + \sin^2 \left( \frac{\pi y}{a}\right)
+ R \left( \cos\left( \frac{\pi x}{a} - \theta \right) + \cos\left( \frac{\pi y}{a} \right) \right)^2
\right],
\label{eq:DWL}
\end{eqnarray}
which is Eq.~(20) of the main text. This optical lattice consists of two square optical lattices and $a$ represents the lattice spacing of the one with the shorter period.
We assume that $V_0 < 0$ for atoms while $V_0>0$ for molecules.
In Fig.~3 of the main text, we show the spatial profile of this potential for $V_0 < 0$, $R=0.59$, and $\theta= \pi/6$. Such an optical lattice is often used to create a double-well optical lattice~\cite{sebby-06,anderlini-07}, whose unit cell is a double well potential, and is advantageous for the proposed scheme in the sense that the band levels of the atomic site have no degeneracy as shown in Fig.~\ref{fig:Esp}, where the eigenenergies of a single atom in the optical lattice at zero quasi-momentum are plotted for $V_0 = -60 E_{\rm R}$, $R=0.59$, and $\theta=\pi/6$. $E_{\rm R}\equiv \frac{\hbar^2\pi^2}{2ma^2}$ denotes the recoil energy.  If there are any degenerate levels, then $\Delta_{\rm min}=0$ such that the condition (16) can not be satisfied.

To evaluate the energy scales appearing in the necessary conditions (15)-(19), let us specifically choose $^6$Li as the fermionic atoms confined in our system. Use of this species in cold-atom experiments is rather standard. Setting the lattice spacing to be a standard value, namely $d=532\,{\rm nm}$, leads to the recoil energy $E_{\rm R} = h\times 29.2 \,{\rm kHz}$. Taking the values of the parameters used in Fig.~\ref{fig:Esp} and setting $N=16$ immediately give $\max(t_i)\sim h\times 0.5 \,{\rm Hz}$, $\Delta_{\rm min}=h\times 66.7 \,{\rm Hz}$, and $\Delta_{\rm max}=h\times 1.96 \,{\rm MHz}$. If we set $|\nu_s|=h \times 10 \,{\rm Hz}$, $n_{\rm ms}=36$, and $\sigma_g/\sigma_{\nu}=0.3$, then $J=h\times 27.2 \,{\rm Hz}$. Hence, the first condition (15) is safely satisfied. We note that with these values of $N$, $|\nu_s|$, $n_{\rm ms}$, and $\sigma_g/\sigma_{\nu}$ the condition (\ref{eq:last-cond}) is safely satisfied as well.

Since the second condition (16) requires the information of the linewidths, let us first estimate $\Gamma_{\rm ms}$. If a PA molecule consists of one electronically ground-state alkali atom and one electronically excited alkali atom, a typical scale of the linewidth is $\Gamma_{\rm ms} \sim 2\pi \times 10 \,{\rm MHz}$~\cite{jones-06}, which is much larger than $|\nu_s|$ and can not be used for the present scheme. In contrast, if a PA molecule consists of two electronically ground-state alkali atoms, $\Gamma_{\rm ms}$ is much smaller in general. A coherent coupling to such an electronically ground-state molecule can be created with use of the two-photon Raman PA techniques~\cite{jones-06}. For instance, in the case of $^{87}{\rm Rb}$ atoms confined in an optical lattice, molecular states with linewidths as narrow as $\Gamma_{\rm ms} \sim 2\pi \times 1\, {\rm kHz}$ have been observed~\cite{rom-04}. In the case of $^{6}{\rm Li}$ atoms, detailed experimental searches for linewidths of electronically ground-state molecules in optical lattices have not been performed. However, it is known at least that Feshbach molecules of $^{6}{\rm Li}$, which correspond to an electronically ground-state molecular state, can have lifetime as long as $10\,{\rm s}$ in the absence of an optical-lattice potential~\cite{jochim-03,kohler-06}, meaning that its linewidth can be as narrow as $\Gamma_{\rm ms} \sim 2\pi \times 0.1 \,{\rm Hz}$.

As for $\Gamma_{\rm PA}$, state-of-art experiments have developed lasers with ultra-narrow linewidth for application to optical-lattice atom clocks such that the linewidth can be as low as $\Gamma_{\rm PA} \sim 2\pi \times 0.1 \,{\rm Hz}$~\cite{nakajima-10, inaba-13, takamoto-15}. Using the electronically ground-state molecules and the state-of-art lasers, the condition (16) can be overcome in principle. Notice, however, that implementation of such narrow linewidths for all PA lasers with many different frequencies has never been realized thus far.

Furthermore, we assume that $V_0 = 2\times10^5\,E_{\rm R}$ for the molecular optical lattice so that $\Delta_{\rm MB}=h\times 10.9 \,{\rm MHz}$. Since the level spacing of the rotational states of a $^6{\rm Li}_2$ molecule is typically $\tilde{\Delta}\sim h\times 100 \,{\rm MHz}$, the third condition (17) is also satisfied. Finally, we estimate that $|U_{s,s'}|\sim 3 \,{\rm MHz}$ under the assumption that the s-wave scattering lengths between two molecules take a typical value $|a_{\rm s}|\sim 100 a_{\rm B}$, where $a_{\rm B}$ denotes the Bohr radius. With this estimation of $|U_{s,s'}|$, the fourth  and fifth conditions (18) and (19) are satisfied. This means that it is in principle possible to create the modified SYK model (11) at least up to $N=16$ by means of the proposed scheme with the specific choices of the optical lattice potential of Eq.~(\ref{eq:DWL}) and the atomic species of $^6$Li.

\end{document}